\documentclass[showpacs,aps,graphicx,twocolumn]{revtex4}
\usepackage{graphicx}

\begin{document}

\title{Purification of  Logic-Qubit Entanglement}

\author{Lan Zhou$^{1,2}$ and Yu-Bo Sheng$^{2}$\footnote{shengyb@njupt.edu.cn} }
 \address{$^1$College of Mathematics \& Physics, Nanjing University of Posts and Telecommunications, Nanjing,
210003, China\\
$^2$Key Lab of Broadband Wireless Communication and Sensor Network
 Technology,Nanjing University of Posts and Telecommunications, Ministry of
 Education, Nanjing, 210003,
 China\\ }

\date{\today }
\begin{abstract}
Recently,  the theoretical work of Fr\"{o}wis and W. D\"{u}r (Phys. Rev. Lett. \textbf{106}, 110402 (2011)) and the experiment of Lu \emph{et al.} (Nat. Photon.  \textbf{8}, 364  (2014)) both showed that the logic-qubit entanglement has its potential application in future quantum communication and quantum network. However, the entanglement will suffer from the noise and decoherence. In this paper,  we will investigate the entanglement purification for logic-qubit entanglement.  We show that both the bit-flip error and phase-flip error in logic-qubit entanglement can be well purified. Moreover,  the bit-flip error and in  physical-qubit entanglement can be completely corrected. The phase-flip error equals to the bit-flip error in logic-qubit entanglement which can also be purified. This EPP may provide some potential applications in future quantum communication and quantum network.
\end{abstract}
 \maketitle

\section{Introduction}
Entanglement plays an important role in quantum information areas. quantum teleportation \cite{teleportation},  quantum key distribution (QKD)\cite{qkd},  quantum secret sharing (QSS) \cite{qss}, quantum secure direct communication (QSDC)\cite{qsdc1,qsdc2},  and  quantum repeaters \cite{repeater1,repeater2}, all need  entanglement. Before starting the quantum communication protocol, the parties should  set up the maximally entanglement channel first. Usually, they create the entanglement locally and distribute the entangled state to the distant locations in fiber or free space.
Noise is one of the main obstacle in entanglement distribution. It will degrade the entanglement. The degraded entanglement will decrease the efficiency of the communication and also make the quantum communication insecure.

Entanglement purification is to distill the high quality entangled stats from the low quality of entangled
states \cite{purification1,addpurification2,purification2,purification3,purification4,purification5,purification6,purification7,purification8,purification9,purification10,
purification11,purification12,purification13,purification14,purification15,addpurification,purification16,purification17,purification18,purification19,purification20,purification21,purification22}. In 1996, Bennett \emph{et al.} proposed the concept of entanglement purification \cite{purification1}. Subsequently, there are many efficient entanglement purification protocols (EPPs) proposed. For example, in 2001, Pan \emph{et al.} described the feasible EPP with linear optics \cite{purification3}. In 2008, Sheng \emph{et al.} described an EPP which can be repeated to obtain a higher fidelity \cite{purification6}. In 2010, the deterministic EPP was also proposed \cite{purification7}. In 2014, the EPP for hyperentanglement was  presented \cite{purification15}. Recent researches showed that the entanglement purification can be used to benefit  the blind quantum computation \cite{addpurification,purification17}. There are also some important EPPs for  solid systems, such as the EPP for spins \cite{purification18}, short chains of atoms \cite{purification21,purification22}, and so on.

The EPPs described above all focus on the entanglement encoded in the physical qubit directly, for existing quantum communication protocols are usually based on the physical-qubit entanglement.  Recently, Fro\"{w}is and W. D\"{u}r investigated a new type of entanglement, named concatenated Greenberger-Horne-Zeilinger (C-GHZ) state \cite{cghz1}. The C-GHZ state can be written as \cite{cghz1,cghz2,cghz3,cghz4,yan,pan,logicbell1,logicbell2,logicbell3,logicconcentration}
\begin{eqnarray}
|\Phi^{\pm}\rangle_{N,M}=\frac{1}{\sqrt{2}}(|GHZ^{+}_{N}\rangle^{\otimes M} \pm |GHZ^{-}_{N}\rangle^{\otimes M}).\label{logic}
\end{eqnarray}
Here $M$ is the number of the logic qubit and $N$ is the number of  physical qubit in each logic qubit. Each logic qubit is a physical GHZ state of the form
\begin{eqnarray}
|GHZ^{\pm}_{N}\rangle=\frac{1}{\sqrt{2}}(|0\rangle^{\otimes N}\pm |1\rangle^{\otimes N}).
\end{eqnarray}
In 2014, Lu \emph{et al.} realized the first experiment of logic-qubit entanglement in linear optics \cite{pan}. In 2015,  Sheng and Zhou described the first logic Bell-state analysis \cite{logicbell1}. They showed that we can perform the logic-qubit entanglement swapping and it is possible to perform the long-distance quantum communication based on  logic-qubit entanglement \cite{logicbell2,logicbell3}.  These theory and experiment researches may provide an important avenue that the large-scale  quantum networks and the quantum communication may be based on logic-qubit entanglement in future.

Though many EPPs were proposed and discussed, none protocol discusses the purification of logic-qubit entanglement. In this paper, we will investigate the first model of  entanglement purification for logic-qubit entanglement. We show that both the  bit-flip error and phase-flip error in logic-qubit entanglement can be well purified.
With the help of controlled-not (CNOT) gate, the EPP of logic-qubit entanglement can be simplified to  the EPP of  physical-qubit entanglement, which can be easily purified in the next step. Moreover,  we also show that if a bit-flip error  occurs in one of a physical-qubit entanglement locally, it can be well corrected. Moreover, the phase-flip error in one of a physical-qubit entanglement equals to the bit-flip error in the logic qubit entanglement, which can be well purified.

This paper is organized as follows: In Sec. II, we explain the purification for logic qubit error. In Sec. III, we describe the purification for physical qubit error. In Sec. IV, we present a discussion and conclusion.

\section{Purification of Logic-qubit error}
Suppose that Alice and Bob share the maximally entangled state $|\Phi^{+}\rangle_{AB}$ of the form

The four logic Bell states can be described as
\begin{eqnarray}
|\Phi^{+}\rangle_{AB}=\frac{1}{\sqrt{2}}(|\phi^{+}\rangle_{A}|\phi^{+}\rangle_{B}+|\phi^{-}\rangle_{A}|\phi^{-}\rangle_{B}).\label{initial1}
\end{eqnarray}
From Eq. (\ref{initial1}), the Bell states $|\phi^{+}\rangle$ and $|\phi^{-}\rangle$ can be regarded as the logic qubit $|\overline{0}\rangle$ and $|\overline{1}\rangle$, respectively. If a bit-flip error occurs on the logic qubit with the probability of $1-F$, $|\Phi^{+}\rangle_{AB}$ will become $|\Psi^{+}\rangle_{AB}$ of the form
\begin{eqnarray}
|\Psi^{+}\rangle_{AB}=\frac{1}{\sqrt{2}}(|\phi^{+}\rangle_{A}|\phi^{-}\rangle_{B}+|\phi^{-}\rangle_{A}|\phi^{+}\rangle_{B}).\label{initial2}
\end{eqnarray}
Here $|\phi^{\pm}\rangle$ and $|\psi^{\pm}\rangle$ are four physical Bell states of the form
\begin{eqnarray}
|\phi^{\pm}\rangle=\frac{1}{\sqrt{2}}(|0\rangle|0\rangle\pm|1\rangle|1\rangle),\nonumber\\
|\psi^{\pm}\rangle=\frac{1}{\sqrt{2}}(|0\rangle|1\rangle\pm|1\rangle|0\rangle),\label{bell}
\end{eqnarray}
with $|0\rangle$ and $|1\rangle$ are the physical qubit, respectively.
$|\Phi^{+}\rangle_{AB}$ essentially is the state with $m=N=2$ in Eq. (\ref{logic}).
The whole mixed state can be described as
\begin{eqnarray}
\rho_{0}=F|\Phi^{+}\rangle_{AB}\langle\Phi^{+}|+(1-F)|\Psi^{+}\rangle_{AB}\langle\Psi^{+}|.\label{mix1}
\end{eqnarray}

\begin{figure}[!h]
\begin{center}
\includegraphics[width=8cm,angle=0]{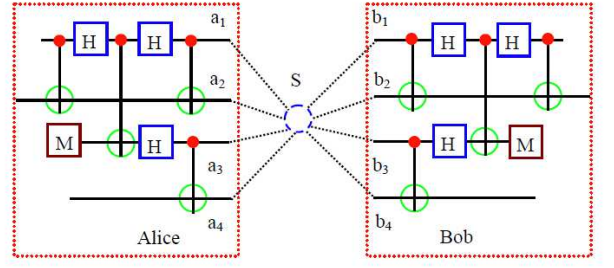}
\caption{Schematic diagram of the purification of logic Bell-state analysis. $H$ represents the Hadamard operation and $M$ represents the measurement in the basis $\{|0\rangle, |1\rangle\}$.}
\end{center}
\end{figure}
As shown in Fig. 1, Alice and Bob share two copies of mixed states, named $\rho_{1}$ and $\rho_{2}$, distributed from
the entanglement source $S$. State $\rho_{1}$ is in the spatial modes $a_{1}$, $a_{2}$, $b_{1}$ and $b_{2}$ and state $\rho_{2}$ is in the spatial modes $a_{3}$, $a_{4}$, $b_{3}$ and $b_{4}$, respectively. The whole system $\rho_{1}\otimes\rho_{2}$ can be described as follows. With the probability of $F^{2}$, it is in the state $|\Phi^{+}\rangle_{A1B1}\otimes|\Phi^{+}\rangle_{A2B2}$. With the equal probability of $F(1-F)$, they are in the states  $|\Phi^{+}\rangle_{A1B1}\otimes|\Psi^{+}\rangle_{A2B2}$ and  $|\Psi^{+}\rangle_{A1B1}\otimes|\Phi^{+}\rangle_{A2B2}$, respectively. With the probability of $(1-F)^{2}$, it is in the state $|\Phi^{+}\rangle_{A1B1}\otimes|\Phi^{+}\rangle_{A2B2}$. Here states $|\Phi^{+}\rangle_{A1B1}$ and $|\Psi^{+}\rangle_{A1B1}$ are the components in $\rho_{1}$ and  $|\Phi^{+}\rangle_{A2B2}$ and $|\Psi^{+}\rangle_{A2B2}$ are the components in $\rho_{2}$, respectively.

We first discuss the item $|\Phi^{+}\rangle_{A1B1}\otimes|\Phi^{+}\rangle_{A2B2}$. It can be written as
\begin{eqnarray}
&&|\Phi^{+}\rangle_{A1B1}\otimes|\Phi^{+}\rangle_{A2B2}\nonumber\\
&=&\frac{1}{\sqrt{2}}(|\phi^{+}\rangle_{A1}|\phi^{+}\rangle_{B1}+|\phi^{-}\rangle_{A1}|\phi^{-}\rangle_{B1})\nonumber\\
&\otimes&\frac{1}{\sqrt{2}}(|\phi^{+}\rangle_{A2}|\phi^{+}\rangle_{B2}+|\phi^{-}\rangle_{A2}|\phi^{-}\rangle_{B2})\nonumber\\
&=&\frac{1}{2}(|\phi^{+}\rangle_{A1}|\phi^{+}\rangle_{A2}|\phi^{+}\rangle_{B1}|\phi^{+}\rangle_{B2}\nonumber\\
&+&|\phi^{+}\rangle_{A1}|\phi^{-}\rangle_{A2}|\phi^{+}\rangle_{B1}|\phi^{-}\rangle_{B2}\nonumber\\
&+&|\phi^{-}\rangle_{A1}|\phi^{+}\rangle_{A2}|\phi^{-}\rangle_{B1}|\phi^{+}\rangle_{B2}\nonumber\\
&+&|\phi^{-}\rangle_{A1}|\phi^{-}\rangle_{A2}|\phi^{-}\rangle_{B1}|\phi^{-}\rangle_{B2}).\label{item1}
\end{eqnarray}
 From Fig. 1, they let all qubits  pass through the controlled-not (CNOT) gate.
  State $|\phi^{+}\rangle_{A1}$ in spatial modes $a_{1}$, $a_{2}$ will become
\begin{eqnarray}
|\phi^{+}\rangle_{A1}&=&\frac{1}{\sqrt{2}}(|0\rangle_{a_{1}}|0\rangle_{a_{2}}+|1\rangle_{a_{1}}|1\rangle_{a_{2}})\nonumber\\
&\rightarrow&\frac{1}{\sqrt{2}}(|0\rangle_{a_{1}}|0\rangle_{a_{2}}+|1\rangle_{a_{1}}|0\rangle_{a_{2}})\nonumber\\
&=&|+\rangle_{a_{1}}|0\rangle_{a_{2}}.
\end{eqnarray}
State $|\psi^{+}\rangle_{A1}$ in spatial modes $a_{1}$, $a_{2}$ will become
\begin{eqnarray}
|\phi^{-}\rangle_{A1}&=&\frac{1}{\sqrt{2}}(|0\rangle_{a_{1}}|0\rangle_{a_{2}}-|1\rangle_{a_{1}}|1\rangle_{a_{2}})\nonumber\\
&\rightarrow&\frac{1}{\sqrt{2}}(|0\rangle_{a_{1}}|0\rangle_{a_{2}}-|1\rangle_{a_{1}}|0\rangle_{a_{2}})\nonumber\\
&=&|-\rangle_{a_{1}}|0\rangle_{a_{2}}.
\end{eqnarray}
Here $|\pm\rangle=\frac{1}{\sqrt{2}} (|0\rangle\pm|1\rangle)$.
After passing through the CNOT gates and  Hadamard gates, with the probability of $F^{2}$, state in Eq. (\ref{item1}) can be evolved as
\begin{eqnarray}
&&|\Phi^{+}\rangle_{A1B1}\otimes|\Phi^{+}\rangle_{A2B2}\nonumber\\
&\rightarrow&\frac{1}{2}(|+\rangle_{a_{1}}|0\rangle_{a_{2}}|+\rangle_{a_{3}}|0\rangle_{a_{4}}|+\rangle_{b_{1}}|0\rangle_{b_{2}}|+\rangle_{b_{3}}|0\rangle_{b_{4}}\nonumber\\
&+&|+\rangle_{a_{1}}|0\rangle_{a_{2}}|-\rangle_{a_{3}}|0\rangle_{a_{4}}|+\rangle_{b_{1}}|0\rangle_{b_{2}}|-\rangle_{b_{3}}|0\rangle_{b_{4}}\nonumber\\
&+&|-\rangle_{a_{1}}|0\rangle_{a_{2}}|+\rangle_{a_{3}}|0\rangle_{a_{4}}|-\rangle_{b_{1}}|0\rangle_{b_{2}}|+\rangle_{b_{3}}|0\rangle_{b_{4}}\nonumber\\
&+&|-\rangle_{a_{1}}|0\rangle_{a_{2}}|-\rangle_{a_{3}}|0\rangle_{a_{4}}|-\rangle_{b_{1}}|0\rangle_{b_{2}}|-\rangle_{b_{3}}|0\rangle_{b_{4}})\nonumber\\
&\rightarrow&|\phi^{+}\rangle_{a_{1}b_{1}}|\phi^{+}\rangle_{a_{3}b_{3}}|0\rangle_{a_{2}}|0\rangle_{b_{2}}|0\rangle_{a_{4}}|0\rangle_{b_{4}}.\label{evolve1}
\end{eqnarray}
Following the same principle, with the probability of $F(1-F)$, state $|\Phi^{+}\rangle_{A1B1}\otimes|\Psi^{+}\rangle_{A2B2}$ can be evolved as
\begin{eqnarray}
&&|\Phi^{+}\rangle_{A1B1}\otimes|\Psi^{+}\rangle_{A2B2}\nonumber\\
&\rightarrow&|\phi^{+}\rangle_{a_{1}b_{1}}|\psi^{+}\rangle_{a_{3}b_{3}}|0\rangle_{a_{2}}|0\rangle_{b_{2}}|0\rangle_{a_{4}}|0\rangle_{b_{4}},\label{evolve2}
\end{eqnarray}
and  state $|\Psi^{+}\rangle_{A1B1}\otimes|\Phi^{+}\rangle_{A2B2}$ can be evolved as
\begin{eqnarray}
&&|\Psi^{+}\rangle_{A1B1}\otimes|\Phi^{+}\rangle_{A2B2}\nonumber\\
&\rightarrow&|\psi^{+}\rangle_{a_{1}b_{1}}|\phi^{+}\rangle_{a_{3}b_{3}}|0\rangle_{a_{2}}|0\rangle_{b_{2}}|0\rangle_{a_{4}}|0\rangle_{b_{4}}.\label{evolve3}
\end{eqnarray}
With the probability of $(1-F)^{2}$, state $|\Psi^{+}\rangle_{A1B1}\otimes|\Psi^{+}\rangle_{A2B2}$ can be evolved as
\begin{eqnarray}
&&|\Psi^{+}\rangle_{A1B1}\otimes|\Psi^{+}\rangle_{A2B2}\nonumber\\
&\rightarrow&|\psi^{+}\rangle_{a_{1}b_{1}}|\psi^{+}\rangle_{a_{3}b_{3}}|0\rangle_{a_{2}}|0\rangle_{b_{2}}|0\rangle_{a_{4}}|0\rangle_{b_{4}}.\label{evolve4}
\end{eqnarray}
Here $|\phi^{+}\rangle_{a_{1}b_{1}}$, $|\psi^{+}\rangle_{a_{1}b_{1}}$,$|\phi^{+}\rangle_{a_{3}b_{3}}$ and $|\psi^{+}\rangle_{a_{3}b_{3}}$ are the physical Bell states as described in Eq. (\ref{bell}) in spatial modes $a_{1}b_{1}$, $a_{3}b_{3}$, respectively.  Interestingly, from Eq. (\ref{evolve1}) to Eq. (\ref{evolve4}), the qubits in spatial modes $a_{2}$, $b_{2}$, $a_{4}$ and $b_{4}$ disentangle with the other qubits. The purification of logic Bell state can be transformed to the purification of the physical Bell state in spatial modes  $a_{1}$, $b_{1}$, $a_{3}$ and $b_{3}$. Briefly speaking, as shown in Fig. 1, they let the qubits in $a_{1}$, $b_{1}$, $a_{3}$ and $b_{3}$ pass through the CNOT gate in a second time. The CNOT gate will make the state \cite{purification1}
\begin{eqnarray}
|\phi^{+}\rangle_{a_{1}b_{1}}|\phi^{+}\rangle_{a_{3}b_{3}}\rightarrow|\phi^{+}\rangle_{a_{1}b_{1}}|\phi^{+}\rangle_{a_{3}b_{3}},\nonumber\\
|\phi^{+}\rangle_{a_{1}b_{1}}|\psi^{+}\rangle_{a_{3}b_{3}}\rightarrow|\phi^{+}\rangle_{a_{1}b_{1}}|\psi^{+}\rangle_{a_{3}b_{3}},\nonumber\\
|\psi^{+}\rangle_{a_{1}b_{1}}|\phi^{+}\rangle_{a_{3}b_{3}}\rightarrow|\psi^{+}\rangle_{a_{1}b_{1}}|\psi^{+}\rangle_{a_{3}b_{3}},\nonumber\\
|\psi^{+}\rangle_{a_{1}b_{1}}|\psi^{+}\rangle_{a_{3}b_{3}}\rightarrow|\psi^{+}\rangle_{a_{1}b_{1}}|\phi^{+}\rangle_{a_{3}b_{3}}.\label{cont}
\end{eqnarray}
Subsequently, Alice and Bob measure their qubits in spatial modes $a_{3}$ and $b_{3}$ in $\{0, 1\}$ basis, respectively. With classical communication, if the measurement results are the same, both $0$, or $1$, the purification is successful. Otherwise, if the measurement results are different, the purification is a failure. From Eq. (\ref{cont}), if it is successful, they will obtain $|\phi^{+}\rangle_{a_{1}b_{1}}$, with the probability of $F^{2}$, and $|\psi^{+}\rangle_{a_{1}b_{1}}$ will the probability of $(1-F)^{2}$.  In this way, they obtain a high fidelity of mixed state
\begin{eqnarray}
\rho_{a_{1}b_{1}}=F'|\phi^{+}\rangle_{a_{1}b_{1}}\langle\phi^{+}|+(1-F')|\psi^{+}\rangle_{a_{1}b_{1}}\langle\psi^{+}|.\label{newmix1}
\end{eqnarray}
Here
\begin{eqnarray}
F'=\frac{F^{2}}{F^{2}+(1-F)^{2}}.
\end{eqnarray}
If $F>\frac{1}{2}$, they can obtain $F'>F$.
State in Eq.(\ref{newmix1}) is the purified physical Bell state. The final step is to recover $\rho_{a_{1}b_{1}}$ to  logic Bell state.
From Fig. 1, they perform the Hadamard operations on the  qubits in spatial modes $a_{1}$ and $b_{1}$ and let four qubits in $a_{1}$, $b_{1}$, $a_{2}$ and $b_{2}$ pass through the CNOT gates, respectively. State $|\phi^{+}\rangle_{a_{1}b_{1}}$ combined with $|0\rangle_{a_{2}}|0\rangle_{b_{2}}$ evolve as
\begin{eqnarray}
&&|\phi^{+}\rangle_{a_{1}b_{1}}|0\rangle_{a_{2}}|0\rangle_{b_{2}}\nonumber\\
&=&\frac{1}{\sqrt{2}}(|0\rangle_{a_{1}}|0\rangle_{b_{1}}+|1\rangle_{a_{1}}|1\rangle_{b_{1}})|0\rangle_{a_{2}}|0\rangle_{b_{2}}\nonumber\\
&\rightarrow&\frac{1}{\sqrt{2}}[\frac{1}{\sqrt{2}}(|0\rangle_{a_{1}}
+|1\rangle_{a_{1}})|0\rangle_{a_{2}}
\otimes\frac{1}{\sqrt{2}}(|0\rangle_{b_{1}}+|1\rangle_{b_{1}})|0\rangle_{b_{2}}\nonumber\\
&+&\frac{1}{\sqrt{2}}(|0\rangle_{a_{1}}
-|1\rangle_{a_{1}})|0\rangle_{a_{2}}\otimes\frac{1}{\sqrt{2}}(|0\rangle_{b_{1}}-|1\rangle_{b_{1}})|0\rangle_{b_{2}}]\nonumber\\
&\rightarrow&\frac{1}{\sqrt{2}}[\frac{1}{\sqrt{2}}(|0\rangle_{a_{1}}|0\rangle_{a_{2}}+|1\rangle_{a_{1}}|1\rangle_{a_{2}})\nonumber\\
&\otimes&\frac{1}{\sqrt{2}}(|0\rangle_{b_{1}}|0\rangle_{b_{2}}+|1\rangle_{b_{1}}|1\rangle_{b_{2}})\nonumber\\
&+&\frac{1}{\sqrt{2}}(|0\rangle_{a_{1}}|0\rangle_{a_{2}}-|1\rangle_{a_{1}}|1\rangle_{a_{2}})\nonumber\\
&\otimes&\frac{1}{\sqrt{2}}(|0\rangle_{b_{1}}|0\rangle_{b_{2}}-|1\rangle_{b_{1}}|1\rangle_{b_{2}})]\nonumber\\
&=&|\Phi^{+}\rangle_{A1B1}.
\end{eqnarray}
Following the same principle, state $|\psi^{+}\rangle_{a_{1}b_{1}}$ combined with $|0\rangle_{a_{2}}|0\rangle_{b_{2}}$ evolve to $|\Psi^{+}\rangle_{A1B1}$.
Finally, they will obtain a new mixed state
\begin{eqnarray}
\rho'_{1}=F'|\Phi^{+}\rangle_{A1B1}\langle\Phi^{+}|+(1-F')|\Psi^{+}\rangle_{A1B1}\langle\Psi^{+}|.\label{mix2}
\end{eqnarray}
In this way, they have completed the purification.

On the other hand, if a phase-flip error occurs, it will make the state in Eq. (\ref{initial1}) become
\begin{eqnarray}
|\Phi^{-}\rangle_{AB}=\frac{1}{\sqrt{2}}(|\phi^{+}\rangle_{A}|\phi^{+}\rangle_{B}-|\phi^{-}\rangle_{A}|\phi^{-}\rangle_{B}).\label{initial3}
\end{eqnarray}
The whole mixed state can be written as
\begin{eqnarray}
\rho'_{2}=F|\Phi^{+}\rangle_{AB}\langle\Phi^{+}|+(1-F)|\Phi^{-}\rangle_{AB}\langle\Phi^{-}|.\label{mix3}
\end{eqnarray}
The mixed state in Eq. (\ref{mix3}) can also be purified with the same principle. Briefly speaking, as shown in Fig. 1, they first choose two copies of the states in  Eq. (\ref{mix3}).
 After the qubits in spatial modes $a_{1}$, $a_{2}$, $b_{1}$, $b_{2}$, $a_{3}$, $a_{3}$, $b_{3}$ and $b_{4}$ passing through the CNOT gates and Hadamard gates, respectively, $|\Phi^{+}\rangle_{A1B1}\otimes|\Phi^{+}\rangle_{A2B2}$ will become $|\phi^{+}\rangle_{a_{1}b_{1}}|\phi^{+}\rangle_{a_{3}b_{3}}|0\rangle_{a_{2}}|0\rangle_{b_{2}}|0\rangle_{a_{4}}|0\rangle_{b_{4}}$, which is shown in Eq. (\ref{evolve1}). State $|\Phi^{+}\rangle_{A1B1}\otimes|\Phi^{-}\rangle_{A2B2}$ will become $|\phi^{+}\rangle_{a_{1}b_{1}}|\phi^{-}\rangle_{a_{3}b_{3}}|0\rangle_{a_{2}}|0\rangle_{b_{2}}|0\rangle_{a_{4}}|0\rangle_{b_{4}}$.
 State $|\Phi^{-}\rangle_{A1B1}\otimes|\Phi^{+}\rangle_{A2B2}$ will become $|\phi^{-}\rangle_{a_{1}b_{1}}|\phi^{+}\rangle_{a_{3}b_{3}}|0\rangle_{a_{2}}|0\rangle_{b_{2}}|0\rangle_{a_{4}}|0\rangle_{b_{4}}$, and state
 $|\Phi^{-}\rangle_{A1B1}\otimes|\Phi^{-}\rangle_{A2B2}$ will become $|\phi^{-}\rangle_{a_{1}b_{1}}|\phi^{-}\rangle_{a_{3}b_{3}}|0\rangle_{a_{2}}|0\rangle_{b_{2}}|0\rangle_{a_{4}}|0\rangle_{b_{4}}$. They only need to add the Hadamard operations on each qubit, which make $|\phi^{+}\rangle$ do not change, and $|\phi^{-}\rangle$  become $|\psi^{+}\rangle$.  They essentially transform the phase-flip error to bit-flip error, which has the same form described above. In this way, the phase-flip error in logic-qubit entanglement can also be purified.

\begin{figure}[!h]
\begin{center}
\includegraphics[width=8cm,angle=0]{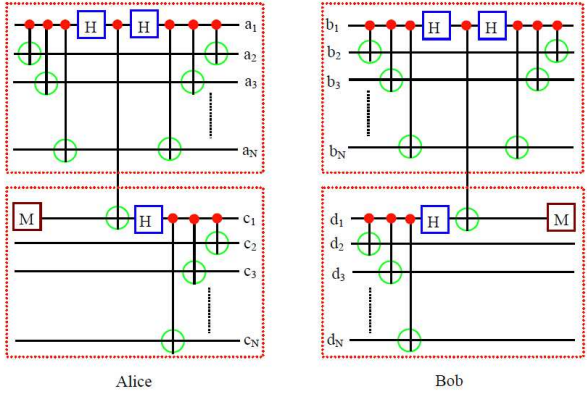}
\caption{Schematic diagram of the EPP with each logic qubit being arbitrary GHZ state. On pair of mixed state $\rho_{ab}$ is in the spatial modes $a_{1}$, $b_{1}$, $a_{2}$, $b_{2}$, $\cdots$, $a_{N}$ and $b_{N}$. The other copy of mixed state $\rho_{cd}$ is in the spatial modes $c_{1}$, $d_{1}$, $c_{2}$, $d_{2}$, $\cdots$, $c_{N}$ and $d_{N}$. }
\end{center}
\end{figure}
So far, we have described the EPP for logic-qubit entanglement. Each logic qubit is encoded in a physical Bell state. It is straightforward to extend
this approach to the logic-qubit entanglement with arbitrary physical GHZ state encoded in a logic qubit. Suppose that Alice and Bob share the state
\begin{eqnarray}
|\Phi_{1}^{+}\rangle_{AB}&=&\frac{1}{\sqrt{2}}(|GHZ_{N}^{+}\rangle_{A}|GHZ_{N}^{+}\rangle_{B}\nonumber\\
&+&|GHZ_{N}^{-}\rangle_{A}|GHZ_{N}^{-}\rangle_{B}).\label{initial4}
\end{eqnarray}
The noise makes the state become
\begin{eqnarray}
\rho_{3}=F|\Phi_{1}^{+}\rangle_{AB}\langle\Phi_{1}^{+}|+(1-F)|\Psi_{1}^{+}\rangle_{AB}\langle\Psi_{1}^{+}|.\label{mix4}
\end{eqnarray}
Here
\begin{eqnarray}
|\Psi_{1}^{+}\rangle_{AB}&=&\frac{1}{\sqrt{2}}(|GHZ_{N}^{+}\rangle_{A}|GHZ_{N}^{-}\rangle_{B}\nonumber\\
&+&|GHZ_{N}^{-}\rangle_{A}|GHZ_{N}^{+}\rangle_{B}).\label{initial5}
\end{eqnarray}
As shown in Fig. 2, they first choose two copies of the mixed states with the same form of $\rho_{3}$. One mixed state $\rho_{ab}$ is in the spatial modes $a_{1}$, $b_{1}$, $a_{2}$, $b_{2}$, $\cdots$, $a_{N}$, $b_{N}$, and the other mixed state  $\rho_{cd}$ is in the mixed state $c_{1}$, $d_{1}$, $c_{2}$, $d_{2}$, $\cdots$, $c_{N}$, $d_{N}$, respectively. We first discuss the mixed state $\rho_{ab}$. After passing through the CNOT gates and Hadamard gates,
with the probability of $F$, state $|\Phi_{1}^{+}\rangle_{ab}$ becomes
\begin{eqnarray}
&&|\Phi_{1}^{+}\rangle_{ab}=\frac{1}{\sqrt{2}}(|GHZ_{N}^{+}\rangle_{a}|GHZ_{N}^{+}\rangle_{b}\nonumber\\
&+&|GHZ_{N}^{-}\rangle_{a}|GHZ_{N}^{-}\rangle_{b})\nonumber\\
&=&\frac{1}{\sqrt{2}}[\frac{1}{\sqrt{2}}(|0\rangle_{a_{1}}|0\rangle_{a_{2}}\cdots|0\rangle_{a_{N}}+|1\rangle_{a_{1}}|1\rangle_{a_{2}}\cdots|1\rangle_{a_{N}})\nonumber\\
&\otimes&\frac{1}{\sqrt{2}}(|0\rangle_{b_{1}}|0\rangle_{b_{2}}\cdots|0\rangle_{b_{N}}+|1\rangle_{b_{1}}|1\rangle_{b_{2}}\cdots|1\rangle_{b_{N}})\nonumber\\
&+&\frac{1}{\sqrt{2}}(|0\rangle_{a_{1}}|0\rangle_{a_{2}}\cdots|0\rangle_{a_{N}}-|1\rangle_{a_{1}}|1\rangle_{a_{2}}\cdots|1\rangle_{a_{N}})\nonumber\\
&\otimes&\frac{1}{\sqrt{2}}(|0\rangle_{b_{1}}|0\rangle_{b_{2}}\cdots|0\rangle_{b_{N}}-|1\rangle_{b_{1}}|1\rangle_{b_{2}}\cdots|1\rangle_{b_{N}})]\nonumber\\
&\rightarrow&\frac{1}{\sqrt{2}}[\frac{1}{\sqrt{2}}(|0\rangle_{a_{1}}|0\rangle_{a_{2}}\cdots|0\rangle_{a_{N}}
+|1\rangle_{a_{1}}|0\rangle_{a_{2}}\cdots|0\rangle_{a_{N}})\nonumber\\
&\otimes&\frac{1}{\sqrt{2}}(|0\rangle_{b_{1}}|0\rangle_{b_{2}}\cdots|0\rangle_{b_{N}}+|1\rangle_{b_{1}}|0\rangle_{b_{2}}\cdots|0\rangle_{b_{N}})\nonumber\\
&+&\frac{1}{\sqrt{2}}(|0\rangle_{a_{1}}|0\rangle_{a_{2}}\cdots|0\rangle_{a_{N}}-|1\rangle_{a_{1}}|0\rangle_{a_{2}}\cdots|0\rangle_{a_{N}})\nonumber\\
&\otimes&\frac{1}{\sqrt{2}}(|0\rangle_{b_{1}}|0\rangle_{b_{2}}\cdots|0\rangle_{b_{N}}-|1\rangle_{b_{1}}|0\rangle_{b_{2}}\cdots|0\rangle_{b_{N}})]\nonumber\\
&\rightarrow&\frac{1}{\sqrt{2}}(|0\rangle_{a_{1}}|0\rangle_{b_{1}}+|1\rangle_{a_{1}}|1\rangle_{b_{1}})\nonumber\\
&\otimes&|0\rangle_{a_{2}}\cdots|0\rangle_{a_{N}}|0\rangle_{b_{2}}\cdots|0\rangle_{b_{N}}.\label{extend1}
\end{eqnarray}
With the same principle, with the probability of $1-F$, state $|\Psi_{1}^{+}\rangle_{ab}$ becomes
\begin{eqnarray}
|\Psi_{1}^{+}\rangle_{ab}&\rightarrow&\frac{1}{\sqrt{2}}(|0\rangle_{a_{1}}|1\rangle_{b_{1}}+|1\rangle_{a_{1}}|0\rangle_{b_{1}})\nonumber\\\label{extend2}
&\otimes&|0\rangle_{a_{2}}\cdots|0\rangle_{a_{N}}|0\rangle_{b_{2}}\cdots|0\rangle_{b_{N}}.
\end{eqnarray}

Similar to Eqs. (\ref{extend1}) and (\ref{extend2}), after passing through the CNOT gates and Hadamard gates, state $\rho_{cd}$ in the spatial modes
 $c_{1}$, $d_{1}$, $c_{2}$, $d_{2}$, $\cdots$, $c_{N}$, $d_{N}$ can also evolve as
 \begin{eqnarray}
|\Phi_{1}^{+}\rangle_{cd}&\rightarrow&\frac{1}{\sqrt{2}}(|0\rangle_{c_{1}}|0\rangle_{d_{1}}+|1\rangle_{c_{1}}|1\rangle_{d_{1}})\nonumber\\
&\otimes&|0\rangle_{c_{2}}\cdots|0\rangle_{c_{N}}|0\rangle_{d_{2}}\cdots|0\rangle_{d_{N}},\label{extend3}
\end{eqnarray}
and
\begin{eqnarray}
|\Psi_{1}^{+}\rangle_{cd}&\rightarrow&\frac{1}{\sqrt{2}}(|0\rangle_{c_{1}}|1\rangle_{d_{1}}+|1\rangle_{c_{1}}|0\rangle_{d_{1}})\nonumber\\
&\otimes&|0\rangle_{c_{2}}\cdots|0\rangle_{c_{N}}|0\rangle_{d_{2}}\cdots|0\rangle_{d_{N}}.\label{extend4}
\end{eqnarray}
Here the subscripts $a$, $b$, $c$ and $d$ are the spatial modes as shown in Fig. 2.  From Eqs. (\ref{extend1}) to  (\ref{extend4}), by choosing two copies of mixed states $\rho_{ab}$  and $\rho_{cd}$, they can be simplified to the purification of the physical Bell state, which can be easily performed, similar to Eqs. (\ref{evolve1}) to (\ref{evolve4}). After they obtaining the purified high fidelity physical Bell state, the last step is also to recover Bell state to arbitrary logic Bell state. They first perform the Hadamard operation on the qubit in $a_{1}$ and $b_{1}$, respectively. Subsequently, they both let the $N$ qubits pass through $N-1$ CNOT gates, respectively. Finally, they can obtain a high fidelity of arbitrary logic-qubit entangled state.

 On the other hand, if a phase-flip error occurs, it makes the state $|\Phi_{1}^{+}\rangle_{AB}$ become $|\Phi_{1}^{-}\rangle_{AB}$, which can be written as
 \begin{eqnarray}
|\Phi_{1}^{-}\rangle_{AB}&=&\frac{1}{\sqrt{2}}(|GHZ_{N}^{+}\rangle_{A}|GHZ_{N}^{+}\rangle_{B}\nonumber\\
&-&|GHZ_{N}^{-}\rangle_{A}|GHZ_{N}^{-}\rangle_{B}).\label{initial5}
\end{eqnarray}
The mixed state can be written as
\begin{eqnarray}
\rho_{4}=F|\Phi_{1}^{+}\rangle_{AB}\langle\Phi_{1}^{+}|+(1-F)|\Phi_{1}^{-}\rangle_{AB}\langle\Phi_{1}^{-}|.\label{mix5}
\end{eqnarray}
Interestingly, after passing through the CNOT gates and Hadamard gates, state $|\Phi_{1}^{-}\rangle_{ab}$ will become
\begin{eqnarray}
|\Phi_{1}^{-}\rangle_{cd}&\rightarrow&\frac{1}{\sqrt{2}}(|0\rangle_{a_{1}}|0\rangle_{b_{1}}-|1\rangle_{a_{1}}|1\rangle_{b_{1}})\nonumber\\
&\otimes&|0\rangle_{a_{2}}\cdots|0\rangle_{a_{N}}|0\rangle_{b_{2}}\cdots|0\rangle_{b_{N}}.\label{extend5}
\end{eqnarray}
From Eqs. (\ref{extend1}), (\ref{mix5}) and (\ref{extend5}), we can find that the phase-flip error in the logic-qubit entanglement can be simplified into
the phase-flip error of the physical-qubit Bell entanglement, which can be transformed to the bit-flip error and be purified in the next step. In this way, they can purify arbitrary logic-qubit entanglement.

\section{Purification of physical-qubit error}
In above section, we showed that the bit-flip error and phase-flip error in the logic-qubit entanglement can be simplified into the  bit-flip error and phase-flip error
in the physical-qubit entanglement, respectively. Subsequently, the errors in the physical-qubit entanglement can be well purified with the similar approach as Refs.\cite{purification1,addpurification2}. Besides the errors in the logic-qubit entanglement, the single physical qubit can also suffer from the error.
For example, as shown in Eq. (\ref{initial1}), if a bit-flip error occurs in one of the physical qubit in the logic-qubit $A$, which makes $|\phi^{+}\rangle_{A}$ become $|\psi^{+}\rangle_{A}$ and $|\phi^{-}\rangle_{A}$ become $|\psi^{-}\rangle_{A}$, respectively. Therefore, if the error occurs, it makes the state $|\Phi^{+}\rangle_{AB}$ become
\begin{eqnarray}
|\Phi^{+}\rangle'_{AB}=\frac{1}{\sqrt{2}}(|\psi^{+}\rangle_{A}|\phi^{+}\rangle_{B}+|\psi^{-}\rangle_{A}|\phi^{-}\rangle_{B}).\label{initial4}
\end{eqnarray}
Compared with Eq. (\ref{initial1}) and Eq. (\ref{initial4}), we find that the error occurs locally. In this way, they only require to choose one copy of the mixed state to perform the error correction. They let the logic-qubit $A$ pass through the CNOT gate. The qubit in $a_{1}$ is the control qubit and the qubit in $a_{2}$ is the target qubit. State in Eq. (\ref{initial1}) will become
\begin{eqnarray}
|\Phi^{+}\rangle_{AB}\rightarrow\frac{1}{\sqrt{2}}(|+\rangle_{a_{1}}|0\rangle_{a_{2}}|\phi^{+}\rangle_{B}+|-\rangle_{a_{1}}|0\rangle_{a_{2}}|\phi^{-}\rangle_{B}),\nonumber\\\label{correction1}
\end{eqnarray}
and state in Eq. (\ref{initial4}) will become
\begin{eqnarray}
|\Phi^{+}\rangle'_{AB}\rightarrow\frac{1}{\sqrt{2}}(|+\rangle_{a_{1}}|1\rangle_{a_{2}}|\phi^{+}\rangle_{B}+|-\rangle_{a_{1}}|1\rangle_{a_{2}}|\phi^{-}\rangle_{B}).\nonumber\\\label{correction2}
\end{eqnarray}
From Eqs. (\ref{correction1}) and (\ref{correction2}), they only need to measure the  physical qubit in  $a_{2}$ in the basis $\{0, 1\}$. If it becomes $|1\rangle$, it means that a bit-flip error occurs. If Alice and Bob exploit the quantum nondemolition  (QND) measurement, which do not destroy the physical qubit, they  are only required to perform a bit-flip operation to correct the bit-flip error. On the other hand, if the measurement is destructive,  they can prepare another physical qubit $|0\rangle$ in $a_{2}$ and perform the CNOT operation with the  physical qubit $a_{1}$ in logic qubit $A$ to recover the whole state to $|\Phi^{+}\rangle_{AB}$. If the bit-flip error occurs on the second logic qubit $B$, they can also completely correct it with the same principle.

If the logic qubit is $N$-particle GHZ state, a bit-flip error on the logic-qubit $A$ will make the state become
\begin{eqnarray}
&&|\Phi_{1}^{+}\rangle'_{AB}=\frac{1}{\sqrt{2}}[\frac{1}{\sqrt{2}}(|0\rangle_{a_{1}}|0\rangle_{a_{2}}\cdots|1\rangle_{a_{N}}\nonumber\\
&+&|1\rangle_{a_{1}}|1\rangle_{a_{2}}\cdots|0\rangle_{a_{N}})\nonumber\\
&\otimes&\frac{1}{\sqrt{2}}(|0\rangle_{b_{1}}|0\rangle_{b_{2}}\cdots|0\rangle_{b_{N}}+|1\rangle_{b_{1}}|1\rangle_{b_{2}}\cdots|1\rangle_{b_{N}})\nonumber\\
&+&\frac{1}{\sqrt{2}}(|0\rangle_{a_{1}}|0\rangle_{a_{2}}\cdots|1\rangle_{a_{N}}-|1\rangle_{a_{1}}|1\rangle_{a_{2}}\cdots|0\rangle_{a_{N}})\nonumber\\
&\otimes&\frac{1}{\sqrt{2}}(|0\rangle_{b_{1}}|0\rangle_{b_{2}}\cdots|0\rangle_{b_{N}}-|1\rangle_{b_{1}}|1\rangle_{b_{2}}\cdots|1\rangle_{b_{N}})].\nonumber\\
\end{eqnarray}
They let the particles $a_{1}$, $a_{2}$, $\cdots$, $a_{N}$ pass through the $N-1$ CNOT gates. In each CNOT gate, particle in $a_{1}$ mode is the control qubit and the other is the target qubit. It makes the state $|\Phi_{1}^{+}\rangle'_{AB}$ become
\begin{eqnarray}
&&|\Phi_{1}^{+}\rangle'_{AB}\rightarrow[(|+\rangle_{a_{1}}|0\rangle_{a_{2}}\cdots|1\rangle_{a_{N}})\nonumber\\
&\otimes&\frac{1}{\sqrt{2}}(|0\rangle_{b_{1}}|0\rangle_{b_{2}}\cdots|0\rangle_{b_{N}}+|1\rangle_{b_{1}}|1\rangle_{b_{2}}\cdots|1\rangle_{b_{N}})\nonumber\\
&+&(|-\rangle_{a_{1}}|0\rangle_{a_{2}}\cdots|1\rangle_{a_{N}})\nonumber\\
&\otimes&\frac{1}{\sqrt{2}}(|0\rangle_{b_{1}}|0\rangle_{b_{2}}\cdots|0\rangle_{b_{N}}-|1\rangle_{b_{1}}|1\rangle_{b_{2}}\cdots|1\rangle_{b_{N}})].\nonumber\\\label{correction3}
\end{eqnarray}
From Eq. (\ref{correction3}), by measuring the physical qubit $a_{N}$ in the basis $\{0, 1\}$ , if it becomes $|1\rangle$, it means that a bit-flip error occurs. Following the same principle, it can be completely corrected.

On the other hand, if a phase-flip error occurs on the logic qubit $A$, which makes $|\phi^{+}\rangle\leftrightarrow|\phi^{-}\rangle$.  The state $|\Phi^{+}\rangle''_{AB}$ with a phase-flip error in logic qubit $A$ can be written as
\begin{eqnarray}
&&|\Phi^{+}\rangle''_{AB}=\frac{1}{\sqrt{2}}(|\phi^{-}\rangle_{A}|\phi^{+}\rangle_{B}+|\phi^{+}\rangle_{A}|\phi^{-}\rangle_{B}).\label{initial5}
\end{eqnarray}
Interestingly, from Eq. (\ref{initial5}), we find that the phase-flip error in the two physical qubits essentially equals to the logic bit-flip error as shown in Eq. (\ref{initial2}). In this way, we have completely explained our EPP.

\section{Discussion and conclusion}
In traditional EPPs for physical-qubit entanglement \cite{purification1,addpurification2}, they should purify two kinds of errors. The one is the bit-flip error and the other is the phase-flip error. Using the CNOT gate, the bit-flip error can be purified directly. The phase-flip error can be transformed to the bit-flip error and be purified in the next step. In our EPP, we show that the logic-qubit entanglement may contain four kinds of errors. The bit-flip error and phase-flip error occur in the logic-qubit entanglement and physical-qubit entanglement, respectively. From our description, the bit-flip error and phase-flip error in logic-qubit entanglement can be simplified to the bit-flip error and phase-flip error in physical-qubit entanglement, which can be purified with the previous approach in the next step. On the other hand, if a bit-flip error occurs in one of logic qubit, the error can be completely corrected locally. Moreover, if a phase-flip error occurs in one of the logic qubit, we find that it equals to the bit-flip error in the logic-qubit entanglement. In this way, all  errors can be purified. In our EPP, the key element to realize the protocol is the CNOT gate.   There are some important progresses in construction of the CNOT gate, which shows that it is possible to realize the deterministic CNOT gate in future experiment \cite{cnot1,cnot2,cnot3,cnot4,cnot5}. For example, with the help  cross-Kerr nonlinearity, Nemoto \emph{et al.} and Lin \emph{et al.} described a near deterministic CNOT gate for polarization photons, respectively \cite{cnot1,cnot2}. Recently, Wei and Deng  designed a  deterministic controlled-not gate on two photonic qubits by two single-photon input-output processes and the readout on an electron-medium spin confined in an optical resonant microcavity \cite{cnot3}. The deterministic CNOT for spins \cite{cnot4}, electron-spin qubits assisted by diamond nitrogen-vacancy centers inside cavities were also discussed \cite{cnot5}.

In conclusion, we have described  the first EPP for logic-qubit entanglement. We first described the purification for both the bit-flip error and phase-flip error  in the logic qubit. The entanglement purification for logic-qubit entanglement can be simplified to the purification of the physical-qubit entanglement, which can be performed in the next step. On the other hand, we also discussed the purification of the errors occurring in the physical-qubit entanglement. The bit-flip error on the physical qubit can be completely corrected locally. The phase-flip error occurs on the physical qubit equals to the bit-flip error on the logic qubit, which can also be well purified. Our EPP is suitable for the case that each logic qubit being arbitrary $N$-particle GHZ state. Our EPP may be useful for future long-distance quantum communication based on logic-qubit entanglement.

\section*{ACKNOWLEDGEMENTS}
This work is supported by the National Natural Science Foundation of
China under Grant No. Grant Nos. 11474168 and 61401222, the Natural Science Foundation of
Jiangsu under Grant No. BK20151502, the Qing Lan Project in Jiangsu Province, and a Project Funded by the Priority
Academic Program Development of Jiangsu Higher Education
Institutions.

\end{document}